\documentclass[aps,prc,preprint,groupedaddress]{revtex4}
\usepackage{multirow}
\usepackage{epsfig}
\usepackage{textcomp}
\usepackage{multirow}
\usepackage{epsfig}
\usepackage{longtable}

\begin{document}
\title{Decay patterns of low-lying $N_{s\bar{s}}$ states to the strangeness channels}
\author{Chun-Sheng An$^{1}$}

\author{Ju-Jun Xie$^{2}$}

\author{Gang Li$^{3}$}\email{gli@mail.qfnu.edu.cn}

\affiliation{1. School of Physical Science and Technology, Southwest University, Chongqing 400715, China\\
2. Institute of Modern Physics, Chinese Academy of Sciences, Lanzhou 730000, China\\
3. School of Physics and Engineering, Qufu Normal University, Qufu 273165, China}

\thispagestyle{empty}

\date{\today}

\begin{abstract}
%
%
Here we investigate the decay patterns of the low-lying hidden strangeness nucleon resonances ($\equiv N_{s\bar{s}}$)
to the strangeness channels by employing the chiral Lagrangian approach, where the $N_{s\bar{s}}$ states are treated as compact pentaquark
states. The $S$-wave decays of these states to the $PB$ (pseudoscalar meson and baryon)
and $VB$ (vector meson and baryon) channels are studied. According to the obtained masses and decay
properties, we find four states, namely, $N_{s\bar{s}}(1874)$ with quantum numbers $I(J^{P})=1/2(1/2^{-})$,
$N_{s\bar{s}}(1885)$ with $I(J^{P})=1/2(3/2^{-})$, $N_{s\bar{s}}(2327)$ with $I(J^{P})=1/2(1/2^{-})$,
and $N_{s\bar{s}}(2252)$ with $I(J^{P})=1/2(3/2^{-})$,
may be associated to the well established nucleon resonances $N^*(1895)$, $N^{*}(1875)$, and the newly predicted
$N^{*}(2355)$ and $N^{*}(2250)$ by CLAS collaboration, respectively. In addition, several other obtained
hidden strangeness nucleon resonances may be expected to be dominant components of the predicted missing
resonances in the literatures.

\end{abstract}


\maketitle

%
\section {Introduction}
Investigation on the hadron exotic states is always one of the most interesting subjects in hadronic physics.
On both theoretical and experimental sides, great efforts have been made to search for
the exotic states, and lots of candidates for exotic mesons have been observed in the
last decade, while most of the previous evidences on the existence of pentaquark states are controversial, till
the observation of two $P_{c}^{+}$ states was announced by the LHCb collaboration in 2015~\cite{Aaij:2015tga}. For recent reviews on the exotic states, see~\cite{Chen:2016qju,Esposito:2016noz,Ali:2017jda,Guo:2017jvc}. One has to notice that
kinds of the hidden charm pentaquark states like $P_{c}^{+}$ was firstly predicted in~\cite{Wu:2010jy}
and systematically studied using the constituent quark model in~\cite{Yuan:2012wz}, and it's argued
that the hidden charm states couple strongly to the charmness channels such as
$J/\psi N$ because of the existence of the $c\bar{c}$ pair. On the other hand, a recent investigation
on the charmness-nucleon sigma term indicates that there should be $\sim0.6\%$ charmness components
in the ground state of nucleon~\cite{Duan:2016rkr}, based on this result, one may also expect existence
of the nucleon resonances with $>99\%$ hidden charm pentaquark components above $4$~GeV.

Analogously, in the strangeness sector, there should be sizable hidden strange pentaquark components in the nucleon and its excitations. In Refs.~\cite{Zou:2005xy,An:2005cj,Riska:2005bh,An:2013daa}, the strangeness contributions to the magnetic
moment, spin and magnetic form factor of the nucleon were investigated, where the results showed that the experimental data for strangeness observable of the nucleon
could be well reproduced by considering the compact strangeness components in the nucleon wave function.
Meanwhile, it is claimed that the sea quark-antiquark pairs contribution to properties of nucleon should be significant~\cite{Bijker:2009up,Santopinto:2010zza,Bijker:2012zza}.
Besides, it is shown that the strangeness
components in $N^*(1535)$ should account for the mass ordering of $N^*(1440)$, $N^*(1535)$, and $\Lambda^*(1405)$~\cite{Liu:2005pm},
and the strong coupling of $N^*(1535)$ to the strangeness channels~\cite{Xie:2007qt,Cao:2009ea,Xie:2017erh}, which are
consistent with the predictions of the chiral perturbation theory~\cite{Kaiser:1995cy,Inoue:2001ip,Bruns:2010sv}.
In addition, it is shown that data for the electromagnetic and strong decays of $N^*(1535)$ can also be well fitted
by taking the strangeness contributions into account~\cite{An:2008xk,An:2009uv,An:2011sb}.

Recently, triggered by the observation of the $P^{+}_{c}$ pentaquark states, the
hidden strange pentaquark states as $\eta^{\prime}N$ and $\phi N$ bound states~\cite{Gao:2017hya}, $K\Sigma^{*}$ and $K^{*}\Sigma$
bound states~\cite{He:2017aps,Huang:2018ehi}, and compact five-quark states~\cite{Li:2017kzj} were investigated. In Ref.~\cite{Xie:2017mbe}, a possible $\phi p$ resonance was investigated in the $\Lambda^+_c \to \pi^0 \phi p$ decay by considering a triangle singularity mechanism, where the obtained $\phi p$ invariant mass distribution agrees with
the existing data. It is very interesting that the obtained hidden strange pentaquark states in those works just lie in
the energy range of several nucleon resonances with negative parity located,
as listed in the Particle Data Group reviews (PDG)~\cite{Pdg2018}.

In Ref.~\cite{Lin:2018kcc}, the strong decay behavior of the hidden strange meson-baryon molecular
was studied. Measurements on the decays of $J/\psi$ and $\psi(2S)$ to $nK_{s}^{0}\bar{\Lambda}$
indicated that the nucleon excitations $N^*(1535)$,
$N^*(1875)$ and $N^*(2120)$ may couple strongly to the $K \Lambda$ channel~\cite{Ablikim:2007ec}.
Moreover, it has been claimed that the nucleon resonances lying at $\sim 2$~GeV contribute
significantly to the $\phi N$ production~\cite{Kiswandhi:2010ub,Ozaki:2009mj}. All the above evidences show us that the hidden strange
pentaquark configuration may be the dominant or notable component in some nucleon resonances.

Consequently, we investigate the decay patterns of the low-lying compact hidden strange pentaquark states ($N_{s\bar{s}}$) using
the chiral Lagrangian approach. Wave functions for the $J^{P}=1/2^{-}$ states
obtained in~\cite{Li:2017kzj} are employed, and those for the $J^{P}=3/2^{-}$ states are derived
using the same method. Limited by the employed approach, only the $S$-wave decay patterns for $N_{s\bar{s}}$ to
strangeness channels are roughly estimated in present work.

The present manuscript is organized as follows.
In Sec.~\ref{sec:Theo}, we briefly present the employed formalism, the numerical results and discussions are
given in Sec.~\ref{sec:num}, finally, Sec.~\ref{sec:conclu} is a concluded summary.

%
%
\section{Formalism}
\label{sec:Theo}

If the final meson is assumed to be emitted by a quark,
to investigate the transitions $N_{s\bar{s}}\rightarrow PB(VB)$, one needs the explicit
Lagrangian for quark-meson-quark interaction, and the wave functions of the $N_{s\bar{s}}$
states. The wave functions for the hidden strange nucleon resonances have been explicitly
studied in~\cite{An:2008tz,Li:2017kzj}, and for the quark-quark-meson interaction, we employ
the chiral Lagrangian approach, which is widely used in Refs.~\cite{Manohar:1983md,Li:1997gd,Zhao:1998fn,Zhao:2002id,Zhong:2009sk}.
Accordingly, in this section, we will briefly introduce the wave functions of the considered
hidden strange nucleon resonances in~\ref{subsec1}, and present a short review about the chiral Lagrangian
approach and apply it to the five-quark system in~\ref{subsec2}.

\subsection{Wave functions of the $N_{s\bar{s}}$ states}
\label{subsec1}

Following Refs.~\cite{An:2008tz,Li:2017kzj}, a general wave function for the low-lying hidden strange
nucleon resonances with $J^{P}=S^{-}$ can be written as
\begin{eqnarray}
\psi_{t,s}^{(i)} &=& \sum_{a,b,c}\sum_{Y,y,T_z,t_z} \sum_{S_{4z},s_z}
C^{[1^4]}_{[31]_a[211]_a} C^{[31]_a}_{[F^{(i)}]_b [S^{(i)}]_c}
[F^{(i)}]_{b,Y,T_z} [S^{(i)}]_{c,S_{4z}}
[211;C]_a \nonumber\\
&&(Y,T,T_z,y,\bar t,t_z|1,1/2,t)
(S_4,S_{4z},1/2,s_z|S,S_{z})\bar\chi_{y,t_z}\bar\xi_{s_z}\varphi_{[5]}\, ,
\label{wfc}
\end{eqnarray}
where $\bar\chi_{y,t_z}$ and $\bar\xi_{s_z}$ represent the
isospinor and spinor of the antiquark, respectively, and
$\varphi_{[5]}$ represents the completely symmetrical orbital wave
function. The first summation involves symbols
$C^{[.]}_{[..][...]}$, which are $S_4$ Clebsch-Gordan (CG) coefficients
for the indicated color ($[211]$), flavor-spin ($[31]$) and flavor
($[F]$) and spin ($[S]$) wave functions of the $qqqq$ system. The
second summation runs over the flavor indices in the $SU(3)$
CG coefficient (with 9 symbols) and the third one runs over the
spin indices in the standard $SU(2)$ CG coefficient.

Explicitly, according to Eq.~(\ref{wfc}), there are five possible pentaquark
configurations that have appropriate symmetry structure and spin $1/2$:
\begin{eqnarray}
|1(1/2^{-})\rangle&=&|qqqs([4]_{X}[211]_{C}[31]_{FS}[211]_{F}[22]_{S})\otimes\bar{s}\nonumber\\
|2(1/2^{-})\rangle&=&|qqqs([4]_{X}[211]_{C}[31]_{FS}[211]_{F}[31]_{S})\otimes\bar{s}\nonumber\\
|3(1/2^{-})\rangle&=&|qqqs([4]_{X}[211]_{C}[31]_{FS}[22]_{F}[31]_{S})\otimes\bar{s}\nonumber\\
|4(1/2^{-})\rangle&=&|qqqs([4]_{X}[211]_{C}[31]_{FS}[31]_{F}[22]_{S})\otimes\bar{s}\nonumber\\
|5(1/2^{-})\rangle&=&|qqqs([4]_{X}[211]_{C}[31]_{FS}[31]_{F}[31]_{S})\otimes\bar{s}\,,
\end{eqnarray}
and three configurations with spin $3/2$:
\begin{eqnarray}
|1(3/2^{-})\rangle&=&|qqqs([4]_{X}[211]_{C}[31]_{FS}[211]_{F}[31]_{S})\otimes\bar{s}\nonumber\\
|2(3/2^{-})\rangle&=&|qqqs([4]_{X}[211]_{C}[31]_{FS}[22]_{F}[31]_{S})\otimes\bar{s}\nonumber\\
|3(3/2^{-})\rangle&=&|qqqs([4]_{X}[211]_{C}[31]_{FS}[31]_{F}[31]_{S})\otimes\bar{s}\,.
\label{3half}
\end{eqnarray}

\begin{figure}[htbp]
\begin{center}
{\includegraphics[scale=0.5]{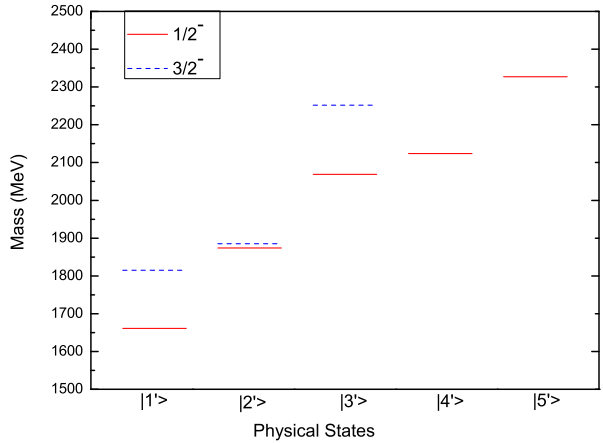}}
\end{center}
{\caption{\footnotesize Spectrum of the low-lying hidden
strange nucleon resonances with $J^{P}=1/2^{-}$ in red (solid) and $J^{P}=3/2^{-}$
in blue (dash), respectively.}
\label{spec}}
\end{figure}

In Ref.~\cite{Li:2017kzj}, the spectrum of the above hidden strange nucleon resonances with spin $1/2$
was studied using the one gluon exchange model (OGE) and Goldstone boson exchange model
(GBE), respectively. One may note that the numerical results obtained in the OGE
model, those are depicted in Fig.~\ref{spec} by solid line in red, should be more reasonable.
So hereafter we will just employ the wave functions obtained in OGE model, the explicit
probability amplitudes for the mixing between $|i(1/2^{-})\rangle$ configurations can be found in~\cite{Li:2017kzj}.
One should note that the numerical results for the energies of the $I(J^{P})=1/2(1/2)^{-}$
states were obtained by using the empirical values for model parameters reported in the literatures, if one
changes values of the coupling strength in OGE model by $\mp10\%$, then the
obtained masses for the five $I(J^{P})=1/2(1/2)^{-}$ $N_{s\bar{s}}$ states will be
$1661\pm51$~MeV, $1874\pm30$~MeV, $2068\pm11$~MeV, $2124\pm6$~MeV and $2327\pm15$~MeV, respectively.
On the other hand, we apply the OGE model in~\cite{Li:2017kzj} to the $3/2^{-}$ $N_{s\bar{s}}$
sector. Using the wave functions as in Eq.~(\ref{3half}) and all the same parameters as in~\cite{Li:2017kzj},
we can get the numerical results shown in Fig.~\ref{spec} by dash line in blue, the explicit values for the energies
of the three obtained physical states are $1815\pm36$~MeV, $1885\pm29$~MeV and $2252\pm7$~MeV, respectively.

Finally, using the empirical values for the coupling strength in OGE model,
the wave functions for the physical states with $J^{P}=3/2^{-}$ are:
\begin{eqnarray}
|1^{\prime}(3/2^{-})\rangle&=&0.989|1(3/2^{-})\rangle-0.151|2(3/2^{-})\rangle+0.008|3(3/2^{-})\rangle\nonumber\\
|2^{\prime}(3/2^{-})\rangle&=&0.150|1(3/2^{-})\rangle+0.985|2(3/2^{-})\rangle+0.082|3(3/2^{-})\rangle\nonumber\\
|3^{\prime}(3/2^{-})\rangle&=&-0.020|1(3/2^{-})\rangle-0.080|2(3/2^{-})\rangle+0.997|3(3/2^{-})\rangle\, .
\label{wfc32}
\end{eqnarray}

\subsection{The chiral Lagrangian approach}
\label{subsec2}

\begin{figure}[t]
\begin{center}
{\hspace{-4cm}\includegraphics[scale=0.4]{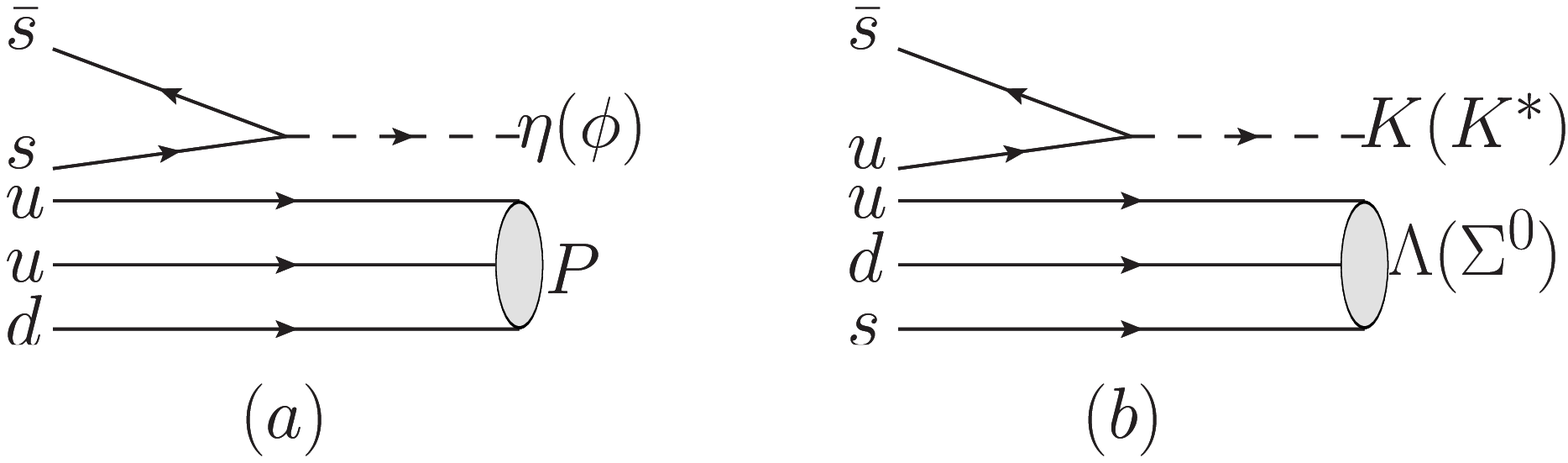}}
\end{center}
\caption{\footnotesize The effective coupling of the $N_{s\bar{s}}$
states to $\eta (\phi) p$ (a), $K (K^{*})\Lambda$ (b) and $K (K^{*})\Sigma$ (c).}
\label{fey}
\end{figure}

In this work, we assume that the final meson couples directly to
a quark and the $\bar{s}$ quark,
namely, a strange quark and the $\bar{s}$ quark annihilate to emit the
$\eta$ or $\phi$ meson, while annihilation of a $u$ or $d$ quark with the $\bar{s}$
quark will emit a $K$ or $K^{*}$ meson, such kind of the quark-meson effective couplings are shown in Fig.~\ref{fey}.

In the chiral Lagrangian approach, the Hamiltonian for quark and pseudoscalar meson
is
\begin{equation}
H_{eff}^{Pqq}=\sum_{j}\bar{\psi}_{j}\gamma_{\mu}^{j}\gamma_{5}^{j}\psi_{j}\partial^{\mu}\phi_{m}\,,
\label{pqq}
\end{equation}
where the summation on $j$ runs over the quark in the initial hadron, $\psi_{j}$ represents the quark field, and
$\phi_{m}$ the pseudoscalar meson field. In the non-relativistic approximation, Eq.~(\ref{pqq}) leads to
the transition operator
\begin{equation}
T_{d}^{Pqq}=\sum_{j}(\frac{\omega_{M}}{E_{f}+M_{f}}\sigma\cdot\vec{P}_{f}+
\frac{\omega_{M}}{E_{i}+M_{i}}\sigma\cdot\vec{P}_{i}-\sigma\cdot\vec{k}_{M}
+\frac{\omega_{M}}{2\mu_{q}}\sigma\cdot\vec{p}_{j})X^{j}_{M}\exp\{-i\vec{k}_{M}\cdot\vec{r}_{j}\}\,,
\end{equation}
for the transition caused by the processes $q\rightarrow q^{\prime}M$, and
\begin{equation}
T_{a}^{Pqq}=\sum_{j}(m_{j}+m_{\bar{q}})
\mathcal{C}^{j}_{XFSC}\bar{\chi}_{z}^{\dag}\mathcal{I}_{2}\chi_{z}^{j}X_{M}^{j}\exp\{-i\vec{k}_{M}\cdot(\vec{r}_{j}+\vec{r}_{\bar{q}})/2\}\,,
\label{tp}
\end{equation}
for the transitions caused by the processes $q\bar{q}\rightarrow M$. Accordingly, Eq.~(\ref{tp}) will
be used in the calculations on the transition matrix elements of $N_{s\bar{s}}\rightarrow PB$
processes. And in~(\ref{tp}), $m_{j}$ and $m_{\bar{q}}$ are the constituent masses of the $jth$ quark and
the antiquark, respectively, $\mathcal{C}^{j}_{XFSC}$ denotes the operator to calculate the orbital, flavor,
spin and color overlap factor between the residual wave function of the pentaquark configuration after the
quark-antiquark annihilation and the wave function of the final baryon, $\bar{\chi}_{z}^{\dag}\mathcal{I}_{2}\chi_{z}^{j}$
is the spin operator for the quark-antiquark annihilation, and $X_{M}^{j}$ is the operator for a pseudoscalar
meson emission, which can be defined as
\begin{eqnarray}
&
X^{j}_{K^{\pm}}=\mp\frac{1}{\sqrt{2}}(\lambda_{4}^{j}\mp i\lambda_{5}^{j}),~X^{j}_{K^{0},\bar{K}^{0}}=
\mp\frac{1}{\sqrt{2}}(\lambda_{6}^{j}\mp i\lambda_{7}^{j}),\nonumber\\
&
X^{j}_{\eta}=cos\theta\lambda_{8}^{j}-sin\theta\sqrt{\frac{2}{3}}\mathcal{I},
~X^{j}_{\eta^\prime}=sin\theta\lambda_{8}^{j}+cos\theta\sqrt{\frac{2}{3}}\mathcal{I}\,,
\label{xm}
\end{eqnarray}
for $K$, $\eta$ and $\eta^{\prime}$ emissions, with $\lambda_{i}^{j}$ the flavor $SU(3)$
Gell-Mann matrices acting on $jth$ quark, and $\mathcal{I}$ is the unit matrix in three-dimensional
space, and $\theta$ is the mixing angle between $\eta_{1}$ and $\eta_{8}$, leading to the physical
$\eta$ and $\eta^{\prime}$, here we take the value $\theta=-23^{\circ}$~\cite{Gobbi:1993au}.

And the Hamiltonian for quark and vector meson reads
\begin{equation}
H_{eff}^{Vqq}=-\sum_{j}\bar{\psi}_{j}(a\gamma_{\mu}^{j}+\frac{ib\sigma_{\mu\nu}k_{M}^{\nu}}{2m_{j}})\phi_{m}^{\mu}\psi_{j}\,,
\label{vqq}
\end{equation}
where $m_{j}$ is the constituent mass of the $jth$ quark, $k_{M}^{\nu}$ represent the four-momentum of the vector meson,
and $\phi_{m}^{\mu}$ the vector meson field, $a$ and $b$ are the vector and tensor coupling constants, respectively.

The quark vector meson coupling in Eq.~(\ref{vqq}) results in the transition operator
\begin{eqnarray}
T^{Vqq}_{d,T}&=&\sum_{j}\left\{i\frac{b^{\prime}}{2m_{j}}\vec{\sigma}_{j}\cdot(\vec{k}_{M}\times\vec{\epsilon})
+\frac{a}{2\mu_{q}}\vec{p}_{j}\cdot\vec{\epsilon}\right\}X_{M}^{j}\exp\{-i\vec{k}_{M}\cdot\vec{r}_{j}\}\,,\\
T^{Vqq}_{d,L}&=&\sum_{j}\frac{aM_{V}}{|\vec{q}|}X_{M}^{j}\exp\{-i\vec{k}_{M}\cdot\vec{r}_{j}\}\,,
\end{eqnarray}
for the transitions caused by the processes $q\rightarrow q^{\prime}M$ with the emitted vector meson
transversely polarized denoted by $T$ and longitudinally polarized denoted by $L$, respectively.
And the operator
\begin{eqnarray}
T^{Vqq}_{a,T}&=&\sum_{j}\left\{a-\frac{m_{j}+m_{\bar{q}}}{2m_{j}}b\right\}\vec{\sigma}\cdot\vec{\epsilon}X_{V}^{j}\exp\{-i\vec{k}_{M}\cdot(\vec{r}_{j}+\vec{r}_{\bar{q}})/2\}\,,\label{tvt}\\
T^{Vqq}_{a,L}&=&\sum_{j}\left\{a-\frac{m_{j}+m_{\bar{q}}}{2m_{j}}b\right\}\frac{E_{V}\vec{\sigma}\cdot\vec{q}}{M_{V}|\vec{q}|}X_{V}^{j}\exp\{-i\vec{k}_{M}\cdot(\vec{r}_{j}+\vec{r}_{\bar{q}})/2\}\label{tvl}\,,
\end{eqnarray}
for the transitions caused by the processes $q\bar{q}\rightarrow M$. Where $X_{V}^{j}$ is the vector meson
emission operator that is defined very similarly to $X_{M}^{j}$ in Eq.~(\ref{xm}), as
\begin{eqnarray}
X^{j}_{K^{*\pm}} &=& \mp\frac{1}{\sqrt{2}}(\lambda_{4}^{j}\mp i\lambda_{5}^{j}),~X^{j}_{K^{*0},\bar{K}^{*0}}=
\mp\frac{1}{\sqrt{2}}(\lambda_{6}^{j}\mp i\lambda_{7}^{j}),\nonumber\\
&&
X^{j}_{\phi}=\frac{\sqrt{2}}{3}\mathcal{I}^{j}-\sqrt{\frac{2}{3}}\lambda_{8}^{j}
\,,
\label{xv}
\end{eqnarray}
for $\rho$, $K^{*}$ and $\bar{K}^{*}$ and $\phi$ emission.
And $E_{V}$ and $M_{V}$ are
the energy and mass of the final meson, and here we have taken
the polarization vector of the final meson to be
\begin{eqnarray}
\epsilon_{\mu}^{L}=\frac{1}{M_{V}}
\pmatrix{ |\vec{k}_{M}| \cr
         E_{V}\frac{\vec{k}_{M}}{|\vec{k}_{M}|}\cr
           }\,,~~~~~~~
\epsilon_{\mu}^{T}=
\pmatrix{ 0 \cr
         \vec{\epsilon}\cr
           }\,,
\end{eqnarray}
with $\vec{\epsilon}(\pm)=1/\sqrt{2}(\mp1,-i,0)^{T}$. And the three-momentum $\vec{k}_{M}$
\begin{equation}
|\vec{k}_{M}|=\frac{\sqrt{[M^{2}_{i}-(M_{f}+m_{M})^{2}][M^{2}_{i}-(M_{f}
-m_{M})^{2}]}}{2M_{i}}\,,
\end{equation}
where $M_{i}$, $M_{f}$ and $m_{M}$ denote masses of the initial $N_{s\bar{s}}$ state, final baryon
and meson, respectively.

%
\section{Results and discussions}
\label{sec:num}

\begin{table*}[htbp]
\caption{The transition matrix elements for the hidden strange pentaquark configurations
with $J^{P}=1/2^{-}$ to PB and VB strangeness channels.
Note that the following common factors are omitted: $(m_{q}+m_{s})\langle\hat{\mathcal{O}}_{X}\rangle$ for $|i(1/2^{-})\rangle\rightarrow PB$
transitions, in addition, $(2\cos\theta+\sqrt{2}\sin\theta)$ for $\eta N$ and
$(2\sin\theta-\sqrt{2}\cos\theta)$ for $\eta^{\prime}N$ decays;
while for $|i(1/2^{-})\rangle\rightarrow VB$ transitions,
a common factor $(a-\frac{m_{q}+m_{s}}{2m_{q}}b)\langle\hat{\mathcal{O}}_{X}\rangle$ for
the transitions with the final meson transversely polarized,
and $(a-\frac{m_{q}+m_{s}}{2m_{q}}b)\frac{E_{V}}{m_{V}}\langle\hat{\mathcal{O}}_{X}\rangle$
the transitions with final meson longitudinally polarized.
\label{1/2}}
\vspace{0.3cm}
\renewcommand\tabcolsep{0.7cm}
\renewcommand{\arraystretch}{0.8}
\begin{tabular}{cccccc}
\hline\hline

              &  $|1\rangle$  &   $|2\rangle$  &  $|3\rangle$       &  $|4\rangle$  &  $|5\rangle$    \\
\hline
$\eta N$      &  $\sqrt{3}/3$ &   $-1$         &  $-\sqrt{2/3}$     &  $\sqrt{3}/3$ & $\sqrt{3}/3$    \\

$K\Lambda$    &  $-1/\sqrt{3}$ &  $1$          &  $\sqrt{6}$        &  $\sqrt{3}$   & $\sqrt{3}$      \\

$K\Sigma$     &  $-\sqrt{3}$   &   $3$         &  $-\sqrt{6}$       &  $\sqrt{3}/3$ & $\sqrt{3}/3$    \\

$\eta^{\prime}N$ & $\sqrt{3}/3$ &   $-1$         &  $-\sqrt{2/3}$     &  $\sqrt{3}/3$ & $\sqrt{3}/3$    \\

$K^{*}\Lambda^{(T)}$   & $\sqrt{2/3}$ &   $\sqrt{2}/3$         &  $2/\sqrt{3}$                  &    $-\sqrt{6}$           &    $\sqrt{2/3}$             \\

$K^{*}\Lambda^{(L)}$  &  $1/\sqrt{3}$ &  $1/3$          &      $\sqrt{2/3}$              &       $-\sqrt{3}$        &       $1/\sqrt{3}$        \\

$K^{*}\Sigma^{(T)}$   & $\sqrt{6}$ &     $\sqrt{2}$       &    $-2/\sqrt{3}$                &     $-\sqrt{2/3}$          &    $\sqrt{6}/9$             \\

$K^{*}\Sigma^{(L)}$  &  $\sqrt{3}$ &     $1$       &      $-\sqrt{2/3}$              &       $-1/\sqrt{3}$        &    $\sqrt{3}/9$           \\

$\phi N^{(T)}$   & $2$ &    $2/\sqrt{3}$        &        $2\sqrt{2}/3$            &      $2$         &       $-2/3$          \\

$\phi N^{(L)}$  &  $\sqrt{2}$ &    $\sqrt{2/3}$        &      $2/3$              &   $\sqrt{2}$            &     $-\sqrt{2}/3$          \\

\hline\hline
\end{tabular}
\end{table*}

\begin{table*}[htbp]
\caption{The transition matrix elements for $|i(3/2^{-})\rangle\rightarrow VB$.
Common factors: for
$N_{s\bar{s}}\rightarrow VB$ decays, $(a-\frac{m_{q}+m_{s}}{2m_{q}}b)\langle\hat{\mathcal{O}}_{X}\rangle$
for the transitions with final meson transversely polarized,
and $(a-\frac{m_{q}+m_{s}}{2m_{q}}b)\frac{E_{V}}{m_{V}}\langle\hat{\mathcal{O}}_{X}\rangle$
for the transitions with final meson
longitudinally polarized.
\label{3/2}}
\vspace{0.3cm}
\renewcommand\tabcolsep{0.7cm}
\renewcommand{\arraystretch}{0.8}
\begin{tabular}{cccc}
\hline\hline

              &  $|1\rangle$  &   $|2\rangle$  &  $|3\rangle$     \\
\hline

$K^{*}\Lambda^{(T)}$   & $-2/\sqrt{3}$ &   $-2/3$         &  $-2\sqrt{2/3}$                      \\

$K^{*}\Lambda^{(L)}$  &  $2\sqrt{2/3}$ &  $2\sqrt{2}/3$          &      $4/\sqrt{3}$            \\

$K^{*}\Sigma^{(T)}$   & $-2\sqrt{3}$ &     $-2$       &    $2\sqrt{2/3}$                     \\

$K^{*}\Sigma^{(L)}$  &  $2\sqrt{6}$ &     $2\sqrt{2}$       &      $-4/\sqrt{3}$                \\

$\phi N^{(T)}$   & $-2\sqrt{2}$ &    $-2\sqrt{2/3}$        &        $-4/3$              \\

$\phi N^{(L)}$  &  $4$ &    $4/\sqrt{3}$        &      $4\sqrt{2}/3$                 \\

\hline\hline
\end{tabular}
\end{table*}

Taking the wave functions of the pentaquark configurations $|i(1/2^{-})\rangle$ and $|i(3/2^{-})\rangle$
and the transition operators given in Sec.~\ref{sec:Theo}, we obtain the S-wave transition
elements for the configurations with $J^{P}=1/2^{-}$ in Table~\ref{1/2} and those for configurations
with $J^{P}=3/2^{-}$ in Table~\ref{3/2}, respectively. Note that in the common factors given in the
captions of these tables, $m_{q}=m_{s}$ applies for the transitions with final states $\eta N$, $\eta^{\prime}N$ and
$\phi N$, and $m_{q}=m$ for the transitions with all the other final states, and $\langle\hat{\mathcal{O}}_{X}\rangle$
is the orbital matrix element that depends on the three-momentum $\vec{k}_{M}$, which reads
\begin{equation}
\langle\hat{\mathcal{O}}_{X}\rangle\propto\exp\{-3k_{M}^{2}/20\omega^{2}\}\,,
\end{equation}
where $\omega$ is the harmonic oscillator parameter.

To get the numerical results, here we take the explicit empirical values for the model
parameters: $m=340$~MeV and $m_{s}=460$~MeV for the constituent masses of the quarks~\cite{Capstick:2000qj},
$f_{K}=f_{\eta}=160$~MeV and $f_{\eta^{\prime}}=280$~MeV for the decay constants of the
mesons~\cite{Zhong:2009sk,Zhao:2001kk}, $\omega=225$~MeV for the harmonic oscillator parameter~\cite{Li:2017kzj}, $a=-3$ and $b=2$
for the vector and tensor coupling constants~\cite{Zhong:2009sk}, finally, masses of the
final hadrons are taken from PDG~\cite{Pdg2018}. With these values for the parameters, we obtain the numerical
results listed in Table~\ref{res}, where we name the obtained $N_{s\bar{s}}$
pentaquark states according to their masses obtained in this work.
Since the partial decay widths of the $N_{s\bar{s}}$ states depend
on a quark-meson coupling constant which falls in a large range,
in Table~\ref{res}, we only show estimations on the ratios
of the transition matrix element squares
\begin{equation}
\mathcal{F}\equiv\frac{1}{2J_{i}+1}\sum_{J_{iz},J_{fz}}|\langle B,J_{fz}|T|N_{s\bar{s}},J_{iz}\rangle|^{2}\,,
\label{wd}
\end{equation}
where $J_{i}$ and $J_{f}$ are the total angular momenta of the initial and final baryon states, respectively,
and $T$ represent the operators given in Eqs.~(\ref{tp}), (\ref{tvt}) and (\ref{tvl}) for corresponding transitions.
The ratios for the $J^{P}=1/2^{-}$ states listed in Table~\ref{res} are defined as: the obtained $\mathcal{F}$ for
$N_{s\bar{s}}\rightarrow PB(VB)$ transitions over that for the $N_{s\bar{s}}(1661)\rightarrow\eta N$ transition,
while the numerical results for $N_{s\bar{s}}(2252)$ with $J^{P}=3/2^{-}$ are the ratios of the obtained $\mathcal{F}$
for the corresponding channels over that for $N_{s\bar{s}}(2252)\rightarrow K^{*}\Lambda$ channel.

\begin{table*}[ht]
\caption{Numerical results for the ratios of the transition matrix element squares $\mathcal{F}$.
\label{res}}
\vspace{0.3cm}
\renewcommand\tabcolsep{0.35cm}
\renewcommand{\arraystretch}{0.8}
\begin{tabular}{ccccccc}
\hline\hline

                                  &                        &                         &  $J^{P}=1/2^{-}$            &                        &                          &       $J^{P}=3/2^{-}$         \\

                                  \hline

                                  &  $N_{s\bar{s}}(1661)$  &   $N_{s\bar{s}}(1874)$  &  $N_{s\bar{s}}(2069)$       &  $N_{s\bar{s}}(2124)$  &  $N_{s\bar{s}}(2327)$    &       $N_{s\bar{s}}(2252)$     \\
\hline
$\eta N$                          &  $1$                &   $0.19$                 &  $0.10$                      &  $0.03$                 & $\sim0$                  &     $-$      \\

$K\Lambda$                        &  $0.38$                &  $1.48$                 &  $0.08$                      &  $0.05$                 & $\sim0$                   &     $-$       \\

$K\Sigma$                         &  $-$          &   $0.65$                &  $0.11$                      &  $\sim0$               & $\sim0$                   &       $-$     \\

$\eta^{\prime}N$                  & $-$           &  $-$           &  $0.19$                      &  $0.05$                 & $\sim0$                    &     $-$        \\

$K^{*}\Lambda$                    & $-$           & $-$            &  $0.98$                     &    $0.67$              &    $0.47$                &    $1$         \\

$K^{*}\Sigma$                     & $-$           &  $-$           &   $-$              &     $4.05$            &    $0.22$                &     $2.22$      \\

$\phi N$                          & $-$           &  $-$           &        $1.47$               &      $0.58$            &       $0.11$              &      $0.35$      \\

\hline\hline
\end{tabular}
\end{table*}
%
In Table~\ref{res}, the numerical results for the obtained $J^{P}=1/2^{-}$ states
to both $PB$ and $VB$ channels are presented, but for the $J^P=3/2^{-}$ sector, only the results for $N_{s\bar{s}}(2252)\rightarrow VB$
are shown. Since the $N_{s\bar{s}}$ states with $J^{P}=3/2^{-}$ decay into $PB$ channels via $D$-wave,
the partial decay widths of these channels should be much smaller than those of the $VB$ channels.
On the other hands, the two lower $N_{s\bar{s}}$ states with $J^{P}=3/2^{-}$ are below thresholds
of all the studied $VB$ strangeness channels. As we can see in a very recent work~\cite{Lin:2018kcc}, the partial decay
widths of the $K^{*}\Sigma$ and $K\Sigma^{*}$ molecular states have been investigated,
and the numerical results show that the branching ratios for $J^{P}=3/2^{-}$ $N_{s\bar{s}}$
states are very small.

The lowest state $N_{s\bar{s}}(1661)$ lies
at the energy very close to $N^*(1650)$, and $\sim 130$~MeV higher than $N^*(1535)$, one may
expect this state to be sizable components in the two lowest $S_{11}$ states. While we have to
notice that spin symmetry of three-quark component of $S_{11}(1650)$ in the traditional quark model
is expected to be the completely symmetric $[3]_{S}$~\cite{Capstick:2000qj}, which should weaken
the transition between the three-quark component
of $S_{11}(1650)$ and the present obtained $N_{s\bar{s}}(1661)$. In addition, as we can see in Table~\ref{res},
$N_{s\bar{s}}(1661)$ should couple strongly to $\eta N$ and $K\Lambda$ channels, this is consistent with
the large branching ratio of the $N^*(1535)$ resonance to $\eta N$ channel~\cite{Pdg2018},
and the strong coupling of $N^{*}(1535)$ to $K\Lambda$ channel predicted by isobar model~\cite{Liu:2005pm}
and chiral perturbation theory~\cite{Bruns:2010sv}. One may notice that non-vanishing coupling constant
of $N^{*}(1650)$ to $K\Sigma$ channel was reported in a very recent work~\cite{Tiator:2018heh},
while as we can see in Ref.~\cite{An:2011sb}, strong coupling between $N^{*}(1650)$ and $K\Sigma$
can be obtained by taking smaller probabilities of strangeness components in $N^{*}(1650)$ than those in $N^{*}(1535)$.

Another nontrivial result is for the decay pattern of the obtained $N_{s\bar{s}}(1874)$,
which falls in the energy range of the nucleon resonance $N^*(1895)$~\cite{Pdg2018},
considering the uncertainties of the present model, one may expect $N_{s\bar{s}}(1874)$
to be dominant or sizable component in $N^*(1895)$. On the other hand, a very recent measurement
on the $K^{*}\Lambda$ photoproduction showed that $N^{*}(1895)$ should contribute
significantly to the $\gamma p\rightarrow K^{*}\Lambda$ reaction~\cite{Anisovich:2017rpe},
and a partial-wave analysis on the $\gamma p\rightarrow K^{+}\Lambda$ and $\pi^{-}p\rightarrow K^{0}\Lambda$
reactions indicated that $N^{*}(1895)$ was unquestionably required in these processes~\cite{Anisovich:2017ygb},
moreover, the measurement on the $\gamma p\rightarrow\eta p$ and $\gamma p\rightarrow\eta^{\prime}p$
reactions showed strong couplings between $N^{*}(1895)$ and both $\eta p$ and $\eta^{\prime}p$~\cite{Kashevarov:2017kqb}.
All of these evidences are consistent with large strangeness components in the $N^{*}(1895)$
resonance. As we can see in Table~\ref{res}, the obtained $N_{s\bar{s}}(1874)$
in present model may couple strongly to the $K\Lambda$ and $K\Sigma$ channels,
ratio of the coupling constants for the $N_{s\bar{s}}(1874)$ resonance to the $K\Lambda$, $K\Sigma$
and $\eta N$ is obtained to be $|g_{N_{s\bar{s}}K\Lambda}:g_{N_{s\bar{s}}K\Sigma}:g_{N_{s\bar{s}}\eta N}|\sim2.81:1.86:1$,
Accordingly, we can conclude that $N_{s\bar{s}}(1874)$ may be the dominant component of the nucleon resonance $N^*(1895)$.

The obtained $N_{s\bar{s}}(2069)$ and $N_{s\bar{s}}(2124)$ resonances in present model
fall in the energy range of the two-star nucleon resonance $N(2120)$ listed in PDG~\cite{Pdg2018},
whose spin is identified to be $3/2$, although the spin $1/2$ may not be completely excluded.
One can also expect $N_{s\bar{s}}(2069)$ and $N_{s\bar{s}}(2124)$ could be related
to the missing nucleon resonances $N^{*}(2030)/N^{*}(2070)$ and $N^{*}(2145)/N^{*}(2195)$
predicted in~\cite{Capstick:1998uh}, respectively, since all these resonances
are predicted to couple with strangeness channels. Moreover, two $S_{11}$
resonances located at $1846\pm47$~MeV and $2113\pm70$~MeV were predicted by using a dynamical coupled channel approach in~\cite{Chen:2002mn},
obviously, one can also associate the present obtained $N_{s\bar{s}}(2124)$ to the latter one.
About the decay properties of $N_{s\bar{s}}(2069)$ and $N_{s\bar{s}}(2124)$,
as one can see in Table~\ref{res}, these two obtained states seem to couple strongly
to the strangeness $VB$ channels. For instance, if one compare the coupling constants for
$N_{s\bar{s}}(2124)$ to $K^{*}\Sigma$ and $K\Lambda$ channels, the obtained ratio is
$|g_{N_{s\bar{s}}K^{*}\Lambda}:g_{N_{s\bar{s}}K\Lambda}|\sim8.97:1$. Consequently, if one
assumes that the present obtained $N_{s\bar{s}}(2124)$ with $J^{P}=1/2^{-}$ corresponds to a dominant
component in $N(2120)$, significant evidence of this resonance must be shown in the reactions with
$K^{*}\Sigma$ final state, otherwise, this assumption should not be appropriate.
Meanwhile, it has been claimed that the nucleon resonances lying at $\sim2$~GeV may contribute
to $N\phi$ production significantly~\cite{Kiswandhi:2010ub,Ozaki:2009mj},
this seems to coincide with our findings on the $N_{s\bar{s}}(2069)$ pentaquark state.
As we can see in Table~\ref{res}, the ratio of the coupling constants (absolute value) for $N_{s\bar{s}}(2069)$
to $\eta N$, $K^{*}\Lambda$ and $\phi N$ is $\sim1:3.17:3.86$.
Finally, the highest $N_{s\bar{s}}(2327)$, also shows strong coupling to the strangeness $VB$ channels,
but there is no solid experimental evidence on the nucleon resonance with negative parity around or above $2300$~MeV~\cite{Pdg2018},
while one may expect that the $N_{s\bar{s}}(2327)$ state can be associated to the $N^{*}(2355)$ with
$J^{P}=1/2^{-}$ predicted in~\cite{Anisovich:2017rpe}, which should also couple to the $K^{*}\Lambda$
channel.

Next we come to the $3/2^{-}$ $N_{s\bar{s}}$ resonances sector.There are two states
lying at $\sim1850\pm50$~MeV, named as $N_{s\bar{s}}(1815)$ and $N_{s\bar{s}}(1885)$, are obtained in this work.
Since these two obtained states are very close to the nucleon resonance
$N^*(1875)$ with $J^{P}=3/2^{-}$, especially the obtained $N_{s\bar{s}}(1885)$, one may
expect them to take sizable components in $N^*(1875)$. While, as we have discussed above, the present studied
$N_{s\bar{s}}$ with $J^{P}=3/2^{-}$ decay into the $PB$ channels via $D$-wave, hence the branching ratios for this kind of decays should be
small~\cite{Lin:2018kcc}. In addition, all the strangeness pseudoscalar meson and decuplet baryons channels
are above the thresholds of these two obtained hidden strange pentaquark states, so the decay channel
with large branching ratio may be the $\pi\Delta$ channel, although the coupling of $N_{s\bar{s}}$
states to non-strange channels may be not so strong, the phase space for $N_{s\bar{s}}(1885)$ to
$\pi\Delta$ channel is very large. However, the present employed model is not well applicable to
the transitions of $N_{s\bar{s}}$ states to non-strange channels, so we can only give the above
qualitative discussions. The highest obtained $N_{s\bar{s}}$ state $N_{s\bar{s}}(2252)$ lies above $2$~GeV, which couples strongly to $K^{*}\Sigma$
and $K^{*}\Lambda$ channels, but relatively weakly to $\phi N$ channel, one may refer this obtained state
to the predicted $N^{*}(2250)$ in~\cite{Anisovich:2017rpe}.

Finally, as we have discussed above, one has to note the following points before going to the final
conclusions: firstly, the present obtained
results in Table~\ref{res} are only ratios for the square of transition matrix elements of each $N_{s\bar{s}}$
state to the strangeness channels, but not the partial decay widths which depend on the quark-meson coupling
constant; secondly, the present model is not well applicable to the transitions of $N_{s\bar{s}}$ states to
non-strangeness channels, e. g. $N_{s\bar{s}}(1885)\rightarrow\pi\Delta$, which channel may be the one
with largest branching ratio of $N_{s\bar{s}}(1885)$.



\section {Summary and conclusions}
\label{sec:conclu}
In present work, we investigate the strong decay patterns of the low-lying hidden
strange pentaquark states with $I(J^{P})=1/2(1/2^{-})$ and $I(J^{P})=1/2(3/2^{-})$ to the strangeness
channels. Limited by the present employed formalism, the decay patterns for non-strange channels are
not included, here only the $S$-wave decays of the $N_{s\bar{s}}$ states to $PB$ and $VB$ strangeness channels
are roughly estimated. In addition, as a directly extension of one of our previous work, we present the masses
and wave functions of three $N_{s\bar{s}}$ pentaquark states with $I(J^{P})=1/2(3/2^{-})$. One has to note
that the two obtained lower $N_{s\bar{s}}$ states are very close to the nucleon excitation $N^*(1875)$,
and the highest one located closely to the predicted $N^{*}(2250)$ by CLAS collaboration.

In the $1/2^{-}$ sector, the present obtained $N_{s\bar{s}}(1661)$, which is $\sim 100$~MeV higher than the
nucleon resonance $N^*(1535)$, couples strongly to $\eta N$ channel, which is in agreement with the large partial
decay width of $N^*(1535)\rightarrow\eta N$ if we take $N_{s\bar{s}}(1661)$ to be the higher Fock component
in the wave function of $N(1535)$. Note that the probability for $N_{s\bar{s}}(1661)$ in the $N^*(1650)$
resonance may be small, although $N_{s\bar{s}}(1661)$ lies so close to $N^*(1650)$, since the spin structure
should weaken the coupling between the three-quark and the $N_{s\bar{s}}$ components.
The $N_{s\bar{s}}(1874)$ state with $J^{P}=1/2^{-}$ obtained in present model couples very strongly
to $K \Lambda$ channel. This finding seems to coincide to the property of the nucleon resonance $N^*(1895)$, thus, one may expect $N_{s\bar{s}}(1874)$ to be the dominant component of $N^*(1895)$.
For the $N_{s\bar{s}}(2069)$ state, the strangeness channel with largest coupling is $\phi N$, this seems to
be consistent with previous predictions that $N^{*}$ located at $\sim 2$ GeV should contribute significantly
to the $\phi N$ production. And one can also expect the obtained $N_{s\bar{s}}(2069)$ and $N_{s\bar{s}}(2124)$
could be related to the predicted missing resonances by Capstick and Roberts in~\cite{Capstick:1998uh}.
The highest obtained hidden strange state, $N_{s\bar{s}}(2327)$, may be associated to the predicted
$N^{*}(2355)$ resonance~\cite{Anisovich:2017rpe}.

Two of the obtained $N_{s\bar{s}}$ states with $J^{P}=3/2^{-}$, i.e. $N_{s\bar{s}}(1815)$ and $N_{s\bar{s}}(1885)$, which are
very close to the nucleon resonance $N^*(1875)$,
are below the thresholds of all strangeness $VB$ channels. Meanwhile, these states decay into $PB$ channels via $D$-wave,
so these two states may mainly decay into $\pi\Delta$ channel via $S$-wave, since the phase space for this channel is very large. We only present the numerical results
for decay patterns of the $N_{s\bar{s}}(2252)$, which may mainly decay to $K^{*}\Sigma$ and $K^{*}\Lambda$ channels,
this is consistent with the predicted $N^{*}(2250)$ resonance by a very recent measurement on $K^{*}$ photoproduction~\cite{Anisovich:2017rpe},
we could wait for the more precise experimental measurements on the $K^{*}$ productions.

%
\begin{acknowledgments}
We thank the anonymous referee for his/her very constructive comments on the manuscript.
This work is partly supported by the National Natural Science Foundation of China under Grant
Nos. 11675131, 11475227, 11735003, 11675091, and 11835015, the Youth Innovation Promotion Association
CAS No. 2016367.
\end{acknowledgments}
%
%
%
%

%

\begin{thebibliography}{99}
%
%
%

\bibitem{Aaij:2015tga}
  R.~Aaij {\it et al.} [LHCb Collaboration],
  ``Observation of $J/\psi p$ Resonances Consistent with Pentaquark States in $\Lambda_b^0 \to J/\psi K^- p$ Decays,''
  Phys.\ Rev.\ Lett.\  {\bf 115}, 072001 (2015)


\bibitem{Chen:2016qju}
  H.~X.~Chen, W.~Chen, X.~Liu and S.~L.~Zhu,
  ``The hidden-charm pentaquark and tetraquark states,''
  Phys.\ Rept.\  {\bf 639}, 1 (2016)

\bibitem{Esposito:2016noz}
  A.~Esposito, A.~Pilloni and A.~D.~Polosa,
  ``Multiquark Resonances,''
  Phys.\ Rept.\  {\bf 668}, 1 (2016)


\bibitem{Ali:2017jda}
  A.~Ali, J.~S.~Lange and S.~Stone,
  ``Exotics: Heavy Pentaquarks and Tetraquarks,''
  Prog.\ Part.\ Nucl.\ Phys.\  {\bf 97}, 123 (2017)

\bibitem{Guo:2017jvc}
  F.~K.~Guo, C.~Hanhart, U.~G.~Meissner, Q.~Wang, Q.~Zhao and B.~S.~Zou,
  ``Hadronic molecules,''
  Rev.\ Mod.\ Phys.\  {\bf 90}, no. 1, 015004 (2018)

\bibitem{Wu:2010jy}
  J.~J.~Wu, R.~Molina, E.~Oset and B.~S.~Zou,
  ``Prediction of narrow $N^*$ and $\Lambda^*$ resonances with hidden charm above 4 GeV,''
  Phys.\ Rev.\ Lett.\  {\bf 105}, 232001 (2010)

\bibitem{Yuan:2012wz}
  S.~G.~Yuan, K.~W.~Wei, J.~He, H.~S.~Xu and B.~S.~Zou,
  ``Study of $qqqc\bar{c}$ five quark system with three kinds of quark-quark hyperfine interaction,''
  Eur.\ Phys.\ J.\ A {\bf 48}, 61 (2012)

\bibitem{Duan:2016rkr}
  S.~Duan, C.~S.~An and B.~Saghai,
  ``Intrinsic charm content of the nucleon and charmness-nucleon sigma term,''
  Phys.\ Rev.\ D {\bf 93}, no. 11, 114006 (2016)

\bibitem{Zou:2005xy}
  B.~S.~Zou and D.~O.~Riska,
  ``The s anti-s component of the proton and the strangeness magnetic moment,''
  Phys.\ Rev.\ Lett.\  {\bf 95}, 072001 (2005)

\bibitem{An:2005cj}
  C.~S.~An, D.~O.~Riska and B.~S.~Zou,
  ``Strangeness spin, magnetic moment and strangeness configurations of the proton,''
  Phys.\ Rev.\ C {\bf 73}, 035207 (2006)

\bibitem{Riska:2005bh}
  D.~O.~Riska and B.~S.~Zou,
  ``The Strangeness form-factors of the proton,''
  Phys.\ Lett.\ B {\bf 636}, 265 (2006)

\bibitem{An:2013daa}
  C.~S.~An and B.~Saghai,
  ``Strangeness magnetic form factor of the proton in the extended chiral quark model,''
  Phys.\ Rev.\ C {\bf 88}, no. 2, 025206 (2013)

\bibitem{Bijker:2009up}
  R.~Bijker and E.~Santopinto,
  ``Unquenched quark model for baryons: Magnetic moments, spins and orbital angular momenta,''
  Phys.\ Rev.\ C {\bf 80}, 065210 (2009)

\bibitem{Santopinto:2010zza}
  E.~Santopinto and R.~Bijker,
  ``Flavor asymmetry of sea quarks in the unquenched quark model,''
  Phys.\ Rev.\ C {\bf 82}, 062202 (2010).

\bibitem{Bijker:2012zza}
  R.~Bijker, J.~Ferretti and E.~Santopinto,
  ``$s\bar{s}$ sea pair contribution to electromagnetic observables of the proton in the unquenched quark model,''
  Phys.\ Rev.\ C {\bf 85}, 035204 (2012).

\bibitem{Liu:2005pm}
  B.~C.~Liu and B.~S.~Zou,
  ``Mass and K Lambda coupling of N*(1535),''
  Phys.\ Rev.\ Lett.\  {\bf 96}, no. 4, 042002 (2006)

\bibitem{Xie:2007qt}
  J.~J.~Xie, B.~S.~Zou and H.~C.~Chiang,
  ``The Role of N*(1535) in pp $\rightarrow$ pp phi and pi- p $\rightarrow$ n phi reactions,''
  Phys.\ Rev.\ C {\bf 77}, 015206 (2008)

\bibitem{Cao:2009ea}
  X.~Cao, J.~J.~Xie, B.~S.~Zou and H.~S.~Xu,
  ``Evidence of N*(1535) resonance contribution in the pn $\rightarrow$ d phi reaction,''
  Phys.\ Rev.\ C {\bf 80}, 025203 (2009)

\bibitem{Xie:2017erh}
  J.~J.~Xie and L.~S.~Geng,
  ``Role of the $N^*(1535)$ in the $\Lambda^+_c \to \bar{K}^0 \eta p$ decay,''
  Phys.\ Rev.\ D {\bf 96}, no. 5, 054009 (2017)

\bibitem{Kaiser:1995cy}
  N.~Kaiser, P.~B.~Siegel and W.~Weise,
  ``Chiral dynamics and the S11 (1535) nucleon resonance,''
  Phys.\ Lett.\ B {\bf 362}, 23 (1995)

\bibitem{Inoue:2001ip}
  T.~Inoue, E.~Oset and M.~J.~Vicente Vacas,
  ``Chiral unitary approach to S wave meson baryon scattering in the strangeness S = O sector,''
  Phys.\ Rev.\ C {\bf 65}, 035204 (2002)

\bibitem{Bruns:2010sv}
  P.~C.~Bruns, M.~Mai and U.~G.~Meissner,
  ``Chiral dynamics of the S11(1535) and S11(1650) resonances revisited,''
  Phys.\ Lett.\ B {\bf 697}, 254 (2011)

\bibitem{An:2008xk}
  C.~S.~An and B.~S.~Zou,
  ``The Role of the qqqq anti-q components in the electromagnetic transition gamma*N $\rightarrow$ N(1535),''
  Eur.\ Phys.\ J.\ A {\bf 39}, 195 (2009)

\bibitem{An:2009uv}
  C.~S.~An and B.~S.~Zou,
  ``Strong decays of N*(1535) in an extended chiral quark model,''
  Sci.\ China G {\bf 52}, 1452 (2009)

\bibitem{An:2011sb}
  C.~An and B.~Saghai,
  ``Strong decay of low-lying $S_{11}$ and $D_{13}$ nucleon resonances to pseudoscalar mesons and octet baryons,''
  Phys.\ Rev.\ C {\bf 84}, 045204 (2011)

\bibitem{Gao:2017hya}
  H.~Gao, H.~Huang, T.~Liu, J.~Ping, F.~Wang and Z.~Zhao,
  ``Search for a hidden strange baryon-meson bound state from $\phi$ production in a nuclear medium,''
  Phys.\ Rev.\ C {\bf 95}, no. 5, 055202 (2017)

\bibitem{Huang:2018ehi}
  H.~Huang, X.~Zhu and J.~Ping,
  ``$P_{c}$-like pentaquarks in hidden strange sector,''
  Phys.\ Rev.\ D {\bf 97}, no. 9, 094019 (2018)

\bibitem{He:2017aps}
  J.~He,
  ``Nucleon resonances $N(1875)$ and $N(2100)$ as strange partners of LHCb pentaquarks,''
  Phys.\ Rev.\ D {\bf 95}, no. 7, 074031 (2017)



\bibitem{Li:2017kzj}
  H.~Li, Z.~X.~Wu, C.~S.~An and H.~Chen,
  ``Low-lying 1/2$^{-}$ hidden strange pentaquark states in the constituent quark model,''
  Chin.\ Phys.\ C {\bf 41}, no. 12, 124104 (2017).

\bibitem{Xie:2017mbe}
  J.~J.~Xie and F.~K.~Guo,
  ``Triangular singularity and a possible $\phi p$ resonance in the $\Lambda^+_c \to \pi^0 \phi p$ decay,''
  Phys.\ Lett.\ B {\bf 774}, 108 (2017).

\bibitem{Pdg2018}
  M. Tanabashi {\it et al.} [Particle Data Group],
  ``Review of Particle Physics,''
  Phys.\ Rev.\ D {\bf 98}, 030001 (2018).


\bibitem{Lin:2018kcc}
  Y.~H.~Lin, C.~W.~Shen and B.~S.~Zou,
  ``Decay behavior of the strange and beauty partners of $P_c$ hadronic molecules,''
  arXiv:1805.06843 [hep-ph].

\bibitem{Ablikim:2007ec}
  M.~Ablikim {\it et al.} [BES Collaboration],
  ``First observation of J / psi and psi(2S) decaying to n K0(S) anti-Lambda + c.c,''
  Phys.\ Lett.\ B {\bf 659}, 789 (2008)

\bibitem{Kiswandhi:2010ub}
  A.~Kiswandhi, J.~J.~Xie and S.~N.~Yang,
  ``Is the nonmonotonic behavior in the cross section of phi photoproduction near threshold a signature of a resonance?,''
  Phys.\ Lett.\ B {\bf 691}, 214 (2010)

\bibitem{Ozaki:2009mj}
  S.~Ozaki, A.~Hosaka, H.~Nagahiro and O.~Scholten,
  ``A Coupled-channel analysis for phi-photoproduction with Lambda(1520),''
  Phys.\ Rev.\ C {\bf 80}, 035201 (2009)
  Erratum: [Phys.\ Rev.\ C {\bf 81}, 059901 (2010)]

\bibitem{An:2008tz}
  C.~S.~An and D.~O.~Riska,
  ``The Small axial charge of the N(1535) resonance,''
  Eur.\ Phys.\ J.\ A {\bf 37}, 263 (2008)


\bibitem{Manohar:1983md}
  A.~Manohar and H.~Georgi,
  ``Chiral Quarks and the Nonrelativistic Quark Model,''
  Nucl.\ Phys.\ B {\bf 234}, 189 (1984).

\bibitem{Li:1997gd}
  Z.~P.~Li, H.~X.~Ye and M.~H.~Lu,
  ``An Unified approach to pseudoscalar meson photoproductions off nucleons in the quark model,''
  Phys.\ Rev.\ C {\bf 56}, 1099 (1997)

\bibitem{Zhao:1998fn}
  Q.~Zhao, Z.~p.~Li and C.~Bennhold,
  ``Vector meson photoproduction with an effective Lagrangian in the quark model,''
  Phys.\ Rev.\ C {\bf 58}, 2393 (1998)

\bibitem{Zhao:2002id}
  Q.~Zhao, J.~S.~Al-Khalili, Z.~P.~Li and R.~L.~Workman,
  ``Pion photoproduction on the nucleon in the quark model,''
  Phys.\ Rev.\ C {\bf 65}, 065204 (2002)

\bibitem{Zhong:2009sk}
  X.~H.~Zhong and Q.~Zhao,
  ``Strong decays of newly observed D(sJ) states in a constituent quark model with effective Lagrangians,''
  Phys.\ Rev.\ D {\bf 81}, 014031 (2010)

\bibitem{Gobbi:1993au}
  C.~Gobbi, F.~Iachello and D.~Kusnezov,
  ``Strong Decays Of Q Anti-Q Mesons,''
  Phys.\ Rev.\  D {\bf 50}, 2048 (1994)


\bibitem{Capstick:2000qj}
  S.~Capstick and W.~Roberts,
  ``Quark models of baryon masses and decays,''
  Prog.\ Part.\ Nucl.\ Phys.\  {\bf 45}, S241 (2000)

\bibitem{Zhao:2001kk}
  Q.~Zhao,
  ``Eta-prime photoproduction near threshold,''
  Phys.\ Rev.\ C {\bf 63}, 035205 (2001).

\bibitem{Tiator:2018heh}
  L.~Tiator {\it et al.},
  ``Eta and Etaprime Photoproduction on the Nucleon with the Isobar Model EtaMAID2018,''
  arXiv:1807.04525 [nucl-th].

\bibitem{Anisovich:2017rpe}
  A.~V.~Anisovich {\it et al.} [CLAS Collaboration],
  ``Differential cross sections and polarization observables from CLAS $K$* photoproduction and the search for new $N$* states,''
  Phys.\ Lett.\ B {\bf 771}, 142 (2017).

\bibitem{Anisovich:2017ygb}
  A.~V.~Anisovich {\it et al.},
  Eur.\ Phys.\ J.\ A {\bf 53}, no. 12, 242 (2017)

\bibitem{Kashevarov:2017kqb}
  V.~L.~Kashevarov {\it et al.} [A2 Collaboration],
  Phys.\ Rev.\ Lett.\  {\bf 118}, no. 21, 212001 (2017)

\bibitem{Capstick:1998uh}
  S.~Capstick and W.~Roberts,
  Phys.\ Rev.\ D {\bf 58}, 074011 (1998)

\bibitem{Chen:2002mn}
  G.~Y.~Chen, S.~Kamalov, S.~N.~Yang, D.~Drechsel and L.~Tiator,
  Nucl.\ Phys.\ A {\bf 723}, 447 (2003)



%
%
%
\end{thebibliography}
\end{document}